\documentclass{optica-article}

\journal{opticajournal} % for journals or Optica Open

\articletype{Research Article}

\usepackage{lineno}
\usepackage{orcidlink}
\usepackage{caption}
\usepackage{subcaption}
\usepackage{hyperref}
\usepackage[Export]{adjustbox}

\newcommand{\p}[1]{\ensuremath{\left( #1 \right)}}

\usepackage[normalem]{ulem}
\usepackage{cancel}
\begin{document}

\title{Rejection of wavefront aberrations in an atomic gradiometer}

\author{
    \orcidlink{0009-0008-4264-7511} Louis Pagot, 
    \orcidlink{0000-0002-4746-2400} S\'ebastien Merlet, 
    \orcidlink{0000-0001-5293-2780} Leonid A. Sidorenkov, 
    and \orcidlink{0000-0003-0659-5028} Franck Pereira Dos Santos\authormark{*}
}

\address{LTE, Observatoire de Paris, Université PSL, Sorbonne Université, Université de Lille, LNE, CNRS \\ 61 Avenue de l'Observatoire, 75014 Paris, France}

\email{\authormark{*}franck.pereira@obspm.fr}

\begin{abstract*} 
One of the main residual limitations of inertial sensors based on atom interferometry stems from laser beam distortions, which cause parasitic phase shifts and non-homogeneous matter-light couplings. Here we present numerical simulations, accompanied by analytical calculations, which quantify the impact of these effects in a cold atom gradiometer. We demonstrate that the propagation of interferometric laser beam aberrations, combined with initial asymmetry and significant time-of-flight expansion of the the two atomic sources, limit the common-mode rejection of phase noise in a differential configuration. The resulting deviations in gravitational acceleration and its gradient are within reach of current experimental devices. Our study allows us to evaluate the surface quality requirements for retroreflective optics in cold-atom gradiometers of various baselines, and can be extended to other sensors based on different interferometer geometries.  
\end{abstract*}

%%%%%%%%%%%%%%%%%%%%%%%%%%  body  %%%%%%%%%%%%%%%%%%%%%%%%%%

%%%%%%%%%%%%%%%%%%%%%%%%%%%%%%%%%%%%%%%%%%%%%%%%%%%%%%%%%%%%%%%%%%%%
%%%%%%%%%%%%%%%%%%%%%%%%%%  Introduction  %%%%%%%%%%%%%%%%%%%%%%%%%%
\section{Introduction}\label{Intro}
Since atom interferometry has been used to determine inertial quantities \cite{riehleOpticalRamseySpectroscopy1991, kasevichAtomicInterferometry1991}, this measurement principle has been applied to various areas of physics, as it allows for accurate and sensitive measurements \cite{Hudson2011, CladeFineStructureCte2019, antoni-micollierDetectingVolcano2022, ArmagnacdeCastanet2024}. % Indeed, a number of 
Many systematic errors and noise sources can be eliminated in atomic interferometers by performing multiple measurements \cite{louchet-chauvetInfluenceTransverseMotion2011}. For example, the $k$-reversal technique allows to eliminate all systematics that are independent of the direction of the effective wave vector. Moreover, since atomic accelerometers, and gravimeters in particular, are limited in sensitivity by the vibrations of the retroreflection  mirror, sensors based on a differential configuration, such as gradiometers, are marginally affected by vibrations that appear as a common-mode noise source\cite{SnaddenMmtTzz1998, CaldaniSimultaneousGraviGradio2019, LyuCompactGraviGradio2022, JanvierCompactDiffGravi2022}. 

The differential configuration also makes it possible to suppress systematic errors common to both interferometers, such as those related to delays between pulses, coupling inhomogeneities due to the finite width of the laser beam, or even Coriolis accelerations and wavefront aberrations, provided that the two atomic clouds have the same kinematic parameters. Thanks to the rejection inherent in this configuration, differential atomic interferometry is well suited to cutting-edge experiments. For example, beyond their direct application in geosciences, gradiometers are used to determine Newton's gravitational constant $\mathcal{G}$ \cite{FixlerAIMmtNewtonianCte2007, RosiPrecisionMmtNewtonianCte2014}. Other examples include long baseline atomic interferometers currently being developed for dark matter searches, as gravitational wave detectors \cite{balaz2025longbaselineatominterferometry}, and to test the universality of free fall \cite{HartwigTestingUFF2015}.

However, rejection based on this differential configuration is not necessarily perfect. The asymmetry of atomic sources and the evolution of the laser field during propagation will lead to residual contributions from the Coriolis effect on the one hand, and from wavefront aberrations on the other, which are the main sources of uncertainty in current experiments \cite{louchet-chauvetInfluenceTransverseMotion2011, LanInfluenceOfCoriolis2012, karcherImprovingAccuracyAtom2018,  LyuCompactGraviGradio2022, LuoEvaluatingEffectWA2025}. Due to their predominant contribution to the accuracy budget of the sensors, these wavefront aberrations have been the subject of numerous studies. In parallel with the development of theoretical frameworks \cite{BadeObservationEPR2018, cervantesEffectAperture2024}, experimental techniques are being elaborated and implemented to measure the effect of these aberrations \textit{in situ} \cite{SeckmeyerPhaseReconstruction2025, junca2025wavefrontmappingabsoluteatom, gaudout2025probingspatialdistributionkvectors}. In some cases, the introduction of controlled distortions and the quantification of their impact with an atomic interferometer itself, have made it possible to efficiently correct the associated systematic errors \cite{schkolnikEffectWavefrontAberrations2015, zhouObservingEffectWavefrontAberrations2016, trimecheActiveControlLaser2017, BadeObservationEPR2018, XuInSituMmtWavefront2024}.

In this article, we examine the impact of laser beam wavefront aberrations in a differential gradiometric configuration, based on numerical simulations and a theoretical framework previously developed to study the influence of optical aberrations on the accuracy of an atomic gravimeter \cite{pagotInfluenceOpticalAberrations2025}. We evaluate the phase residuals of these aberrations due to their finite rejection in a regime where the atomic cloud expands up to a size of the order of the laser beam width. Within this limit, we derive analytical formulas for the curvature of the Gaussian beam and for the distortions introduced by mirror surface defects, described by Zernike polynomials, which allows us to study the dependence of the bias on various experimental parameters. In addition, typical bias ranges are estimated in the presence of fluctuations in the initial transverse positions and velocities of the two atomic clouds, and in the specific cases of a $1$-meter-long baseline gradiometer \cite{CaldaniSimultaneousGraviGradio2019} and of a state-of-the-art $10$-meter-high experiment \cite{HartwigTestingUFF2015, AsenbaumPhaseShiftSpacetimeCurvature2017} 

%%%%%%%%%%%%%%%%%%%%%%%%%%%%%%%%%%%%%%%%%%%%%%%%%%%%%%%%%%%%%%%%%%%%%%%%%%%%%
%%%%%%%%%%%%%%%%%%%%%%%%%%  Gradiometer apparatus  %%%%%%%%%%%%%%%%%%%%%%%%%%
\section{Gradiometer apparatus}\label{System}
\begin{figure}[h]
    \centering
    \rotatebox{0}{\adjincludegraphics[height=8cm]{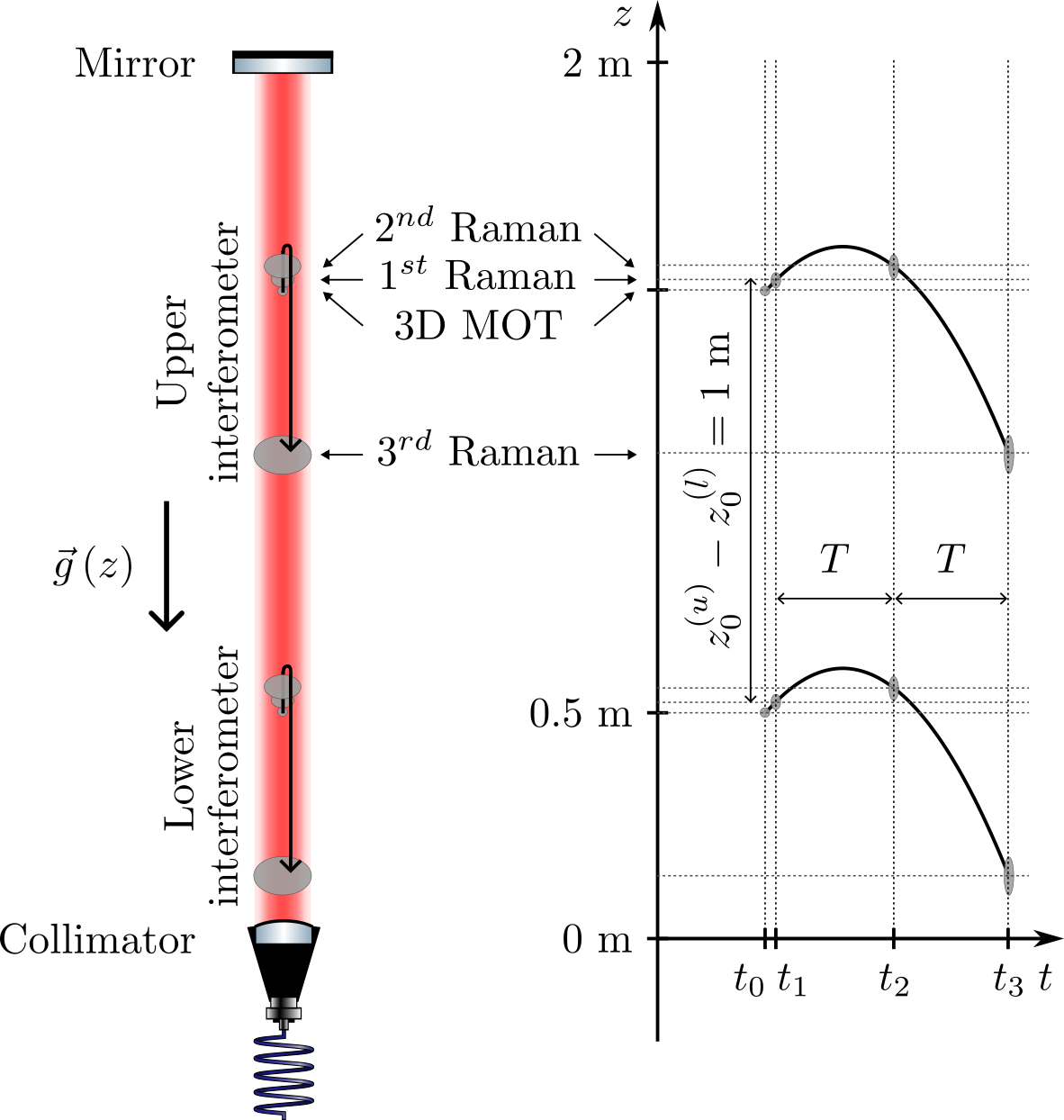}}
    \caption{Gradiometer configuration. The upper and lower atomic clouds are prepared in $3$D~MOTs separated by $1$~m, launched at a velocity $v_z = 1.29$~m~s$^{-1}$, and then an atomic interferometer is performed by successively applying three Raman pulses $\pi/2-\pi-\pi/2$ during free fall. During the time of flight, the clouds expand ballistically and eventually reach the size of the laser beam in the transverse direction.}
  \label{Fig1:Gradiometer}
\end{figure}

The experiment considered here is a $2$-meter-high cold atom gradiometer, illustrated in Figure~\ref{Fig1:Gradiometer}. The output of the laser collimator is placed at height $z=0$~m and the retro-reflecting mirror at the top, at position $z=2$~m. Two $^{87}$Rb atomic clouds are prepared in magneto-optical traps at heights $z^{(l)}=0.50$~m and $z^{(u)}=1.50$~m, then further cooled in optical molasses down to a temperature of $\Theta = 2~\mu$K. At time $t_0=0$~s, they are launched upward at a speed of $1.29$~m~s$^{-1}$. After a Raman velocity selection pulse along the vertical axis, a Mach-Zehnder light pulse atom interferometer, starting at $t_1=15.8$~ms, is performed with a $\pi/2-\pi-\pi/2$ sequence of Raman pulses separated by a time interval $T$ of typically $200$~ms. In this case, the atomic clouds reach their apogee between the first and second pulses. The stimulated Raman transitions are performed using two laser beams with a wavelength $\lambda \approx 780.2$~nm. They arrive through a common collimator with two orthogonal circular polarizations and a radius at $1/e^2$ intensity of $w_0=5$~mm at position $z=0$~m. A combination of a quarter-wave plate and the mirror located at the top of the experiment allows these polarizations to be exchanged for the reflected beams. During the interferometric sequence, the atoms fall freely in the Earth's gravitational potential modeled by the gravity acceleration $g = -9.81$~m~s$^{-2}$ and the vertical gravity gradient $\partial_z g \equiv T_{zz} = 3.08 \times 10^{-6}$~s$^{-2}$ (or 3080 Eötvös, or 3080 E) at $z=0$~m. To compensate for the Doppler effect and scan the inteferometric fringes, the frequency difference between the two lasers is chirped at a rate $\alpha \approx 2\pi \times 25$~MHz~s$^{-1}$. Up to second order in the vertical gravity gradient \cite{APetersHighPrecision2001, hogan2008lightpulseatominterferometry}, the interferometric phase shift $\Delta \phi^{(j)}$ for the lower $\p{j=l}$ and the upper $\p{j=u}$ cloud in the ideal case of a plane wave laser beam is
\begin{equation}\label{Eq:PhaseIdeal}
    \begin{split}
        \Delta \phi^{(j)} = & \thinspace (k_\text{eff} g - \alpha) T\thinspace^2 - k_\text{eff} T\thinspace^2 \thinspace T_{zz} \p{z_0^{(j)} + \overline{v}^{(j)}_0 T - \frac{7}{12} g T\thinspace^2} \\
        & - k_\text{eff} T\thinspace^4 \thinspace T_{zz}^{\thinspace 2} \p{\frac{7}{12}z_0^{(j)} + \frac{1}{4} \overline{v}^{(j)}_0 T + \frac{1}{2} v_\text{rec} T - \frac{31}{360} g T\thinspace^2 }.
    \end{split}
\end{equation}
$k_\text{eff} \approx 2 k = \frac{4\pi}{\lambda}$ is the effective wave vector of the stimulated Raman transition. $z_0^{(j)}$ and $v_0^{(j)}$ are respectively the vertical position and velocity of the atom at the beginning of the interferometer, and $\overline{v}^{(j)}_0 = v_0^{(j)} + v_\text{rec}/2$ is the average velocity of the two paths just after the first interferometric pulse, with $v_\text{rec} = \frac{\hbar k_\text{eff}}{m}$ the two-photon recoil velocity of $^{87}$Rb atoms of mass $m$. The atomic distributions $w$ of each cloud are considered to be independent normal variables with a standard deviation $\sigma_\rho\negthinspace\p{0} = 0.3$~mm in position and $\sigma_v = \sqrt{\frac{k_B\Theta}{m}} = 13.8$~mm~s$^{-1}$ in velocity where $k_B$ is the Boltzmann's constant. After Raman selection along the vertical axis, the width of the vertical velocity distributions of the two clouds is reduced to approximately $6$~mm~s$^{-1}$. In addition, for the following analytical developments, we can consider that the Raman selection only modifies the longitudinal velocity distribution, since the cloud is relatively small compared to the size of the laser beam during this pulse. The atomic cloud distributions after the Raman pulse are of the form
\begin{equation}\label{Eq:DistribCloud}
    w\p{\,\vec{r}, \vec{v}\,} \thinspace dx \thinspace dy \thinspace dz \thinspace dv_x \thinspace dv_y \thinspace dv_z = \frac{e^{-\frac{ x^2 + y^2 }{2 \sigma_\rho^2\negthinspace\p{0}}}}{2\pi \sigma_\rho^2\negthinspace\p{0}} \frac{e^{-\frac{ v_x^2 + v_y^2 }{2 \sigma_v^2}}}{{2\pi \sigma_v^2}} dx \thinspace dy \thinspace dv_x \thinspace dv_y \cdot \eta\negthinspace\p{z, v_z} dz \thinspace dv_z
\end{equation}

Numerically, Monte-Carlo simulations are performed with $25\thinspace000$ to $50\thinspace000$ atoms in each atomic cloud and $11$ to $21$ frequency chirp rates $\alpha$ to scan the interferometer fringes. The parameters obtained by fitting the fringes are eventually averaged over $50$ runs. The transition probabilities of the two interferometers are adjusted to the functions
\begin{equation}\label{Eq:FitFct}
    P^{(j)} = P_m^{(j)} - \frac{C^{(j)}}{2} \cos{\p{\Delta \phi^{(j)} + \delta \phi^{(j)}}}.
\end{equation}
$P_m^{(j)}$ is the mean transition probability, $C^{(j)}$ is the fringe contrast, and $\delta \phi^{(j)}$ is a deviation from the ideal value $\Delta \phi^{(j)}$ given in~\eqref{Eq:PhaseIdeal}. The bias on the vertical gravity gradient and gravity acceleration for each interferometer is calculated as
\begin{equation}\label{Eq:Biases}
    \delta T_{zz} = - \frac{\delta \phi^{(l)} - \delta \phi^{(u)}}{k_\text{eff} T\thinspace^2 \p{z_0^{(l)} - z_0^{(u)}}} \text{ and } \delta g^{(j)} = \frac{\delta \phi^{(j)}}{k_\text{eff} T\thinspace^2}.
\end{equation}

Analytically, the general expression of the transition probability for an atom with initial coordinates $\p{\vec{r}, \vec{v}}$ is
\begin{equation}\label{Eq:GenExpTransitionProba}
    p^{(j)} \negthinspace \p{\,\vec{r}, \vec{v}\,} = p^{(j)}_m \negthinspace \p{\,\vec{r}, \vec{v}\,} - \frac{c^{(j)} \negthinspace \p{\,\vec{r}, \vec{v}\,}}{2} \cos{\p{\Delta \phi^{(j)} + \delta \phi^{(j)} \negthinspace \p{\,\vec{r}, \vec{v}\,} }}.
\end{equation}
The total interferometer phase corresponds to a linear combination of the laser phase acquired by the atom at each pulse: $\phi_1^{(j)} \negthinspace \p{\,\vec{r}, \vec{v}\,} -2 \phi_2^{(j)} \negthinspace \p{\,\vec{r}, \vec{v}\,} + \phi_3^{(j)} \negthinspace \p{\,\vec{r}, \vec{v}\,}$. It is here written as the sum of two terms. $\Delta \phi^{(j)}$: the phase contribution for a plane wave laser beam at the average coordinates of the atomic cloud, which is independent of the atom. $\delta \phi^{(j)} \negthinspace \p{\,\vec{r}, \vec{v}\,}$: is related to the deviation of the wavefront with respect to the plane, and depends on the kinematic parameters of the atom. $c^{(j)} \negthinspace \p{\,\vec{r}, \vec{v}\,}$ is the contrast for this atom. The transition probability for the entire cloud is obtained by integrating expression~\eqref{Eq:GenExpTransitionProba} over the atomic cloud distribution~\eqref{Eq:DistribCloud}. Assuming small phase biases $\left|\delta \phi^{(j)}\p{\vec{r}, \vec{v}}\right| \ll 1$~rad, the approximate expression of contrast \cite{gillotLimitsSymmetryMachZehndertype2016} is
\begin{equation} \label{Eq:ContrastGen}
    C^{(j)} = \int  w\p{\,\vec{r}, \vec{v}\,} \thinspace c^{(j)} \negthinspace \p{\,\vec{r}, \vec{v}\,} \thinspace dx \thinspace dy \thinspace dz \thinspace dv_x \thinspace dv_y \thinspace dv_z,
\end{equation}
and that of phase shift is
\begin{equation} \label{Eq:PhaseShiftGen}
    \delta \phi^{(j)} = \frac{1}{C^{(j)}} \int w\p{\,\vec{r}, \vec{v}\,}  \thinspace c^{(j)} \negthinspace \p{\,\vec{r}, \vec{v}\,} \thinspace \delta \phi^{(j)} \negthinspace \p{\,\vec{r}, \vec{v}\,} \thinspace dx \thinspace dy \thinspace dz \thinspace dv_x \thinspace dv_y \thinspace dv_z.
\end{equation}
Calculating the expression~\eqref{Eq:PhaseShiftGen} requires knowing the expression of the contrast $c^{(j)} \negthinspace \p{\,\vec{r}, \vec{v}\,}$ up to a multiplicative constant. Furthermore, in the simple case where all atoms have a similar interferometric contrast $c^{(j)} \negthinspace \p{\,\vec{r}, \vec{v}\,} \approx \text{cte}$, expression~\eqref{Eq:PhaseShiftGen} simplifies to the phase bias averaged over the atomic cloud distribution: $\delta \phi^{(j)} = \int w\p{\,\vec{r}, \vec{v}\,}  \thinspace \delta \phi^{(j)} \negthinspace \p{\,\vec{r}, \vec{v}\,} \thinspace dx \thinspace dy \thinspace dz \thinspace dv_x \thinspace dv_y \thinspace dv_z$. In this limit of uniform contrast, this relationship can be used to derive the contribution of a Gaussian beam curvature \eqref{Eq:SimpleCurv} \cite{cervantesEffectAperture2024} or that of aberrations described by a Zernike polynomial $Z_n^m$ \cite{pagotInfluenceOpticalAberrations2025}. In the latter case, for a Zernike polynomial defined on a disk of radius $R$, the phase shift is a function of: the amplitude of the aberration, the ratio between the atomic transverse size and the typical transverse length of the aberration $\sigma_\rho/l_{xy}$ with $l_{xy} = R/\p{n+1}$, and the ratio between the distance along the optical axis and the typical propagation length of the aberration $\Delta z / l_z$ where $l_z = 2 k l_{xy}^2$.

%%%%%%%%%%%%%%%%%%%%%%%%%%%%%%%%%%%%%%%%%%%%%%%%%%%%%%%%%%%%%%%%%%%%%%%%%%%%%%%%%
%%%%%%%%%%%%%%%%%%%%%%%%%%  Finite size gaussian beam  %%%%%%%%%%%%%%%%%%%%%%%%%%
\section{Finite size gaussian beam}\label{FiniteSize}
\begin{figure}[h]
    \centering
    \adjincludegraphics[width=\textwidth, trim=0 0 0 1.5cm, clip]{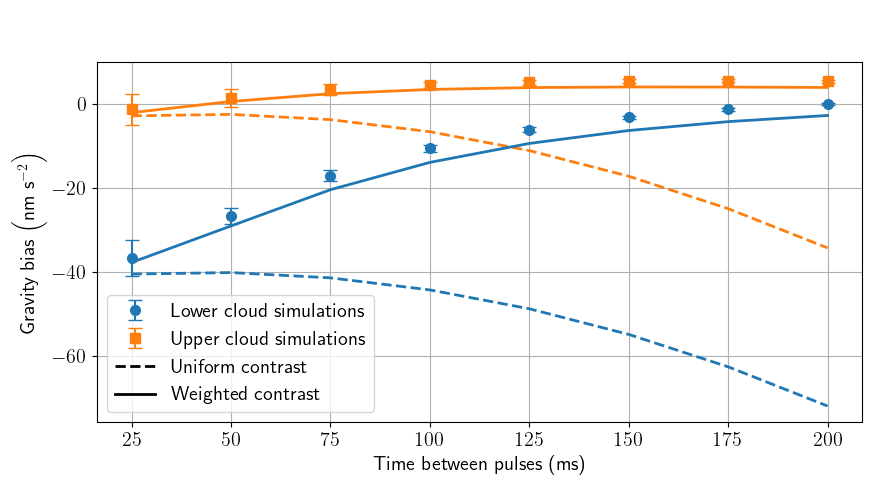}
    \caption{Gravity bias for the two interferometers of the gradiometer. Each data point corresponds to $50$ Monte-Carlo simulations with $400\thinspace000$ atoms to limit the error bars for $T$ close to $25$~ms. Solid (dashed) lines correspond to the analytical expression that (do not) take into account the contrast weights in~\eqref{Eq:PhaseShiftGen}.}
  \label{Fig2:GradiometerFreeFall}
\end{figure}

In the limit of uniform contrast, the phase bias introduced by the curved wavefront of a Gaussian beam \cite{cervantesEffectAperture2024}, for centered clouds, is 
\begin{equation}\label{Eq:SimpleCurv}
    \delta \phi_\text{curv} = 2 \p{\beta_1 \, \sigma_\rho^2\negthinspace\p{t_1} - 2 \beta_2 \, \sigma_\rho^2\negthinspace\p{t_2}+ \beta_3 \, \sigma_\rho^2\negthinspace\p{t_3}}.
\end{equation}
$\beta_i = k \left[ R^{-1}\negthinspace \p{z_i} - R^{-1}\negthinspace \p{z_{i,\text{ref}}}\right]/2$ is proportional to the difference between the inverse of the radius of curvature of the direct and reflected beams. Together with the contribution of the Gouy phase, the gravity biases for the two interferometers of the gradiometer are represented by dashed lines in Figure~\ref{Fig2:GradiometerFreeFall}. These analytical calculations of the biases for the two interferometers diverge from the Monte-Carlo simulations (data points in Figure~\ref{Fig2:GradiometerFreeFall}) as the time interval $T$ between pulses increases, since the assumption of uniform contrast, which requires an atomic cloud smaller than the waist of the laser beam, breaks down. As atomic clouds have normal distributions in transverse position and velocity, their size over time is $\sigma_\rho\negthinspace\p{t} = \sqrt{\sigma^2_\rho\negthinspace\p{0} + \sigma_v^2 t^2}$, which, for a free fall time of $2T \approx 400$~ms, corresponds to $\sigma_\rho = 5.5$~mm, which is similar to the waist $w_0$ of the Gaussian beam. We therefore expect the phase bias to also depend on the ratio between the size of the atomic cloud and the size of the beam: $\sigma_\rho/w_0$ and to reproduce previous results when this ratio is negligible.

To model the contrast loss due to a non-negligible ratio $\sigma_\rho/w_0$ in a simple way, we neglect the impact of inhomogeneities at the first and second pulses, and only take them into account at the final pulse, where the cloud is largest. In addition, since the initial longitudinal distributions of the atomic clouds are independent of the transverse distributions, other effects related to longitudinal coordinates, such as Doppler shift, are assumed to contribute only to a multiplicative constant in the expression of the contrast. Thus, using the analytical expression from \cite{gillotLimitsSymmetryMachZehndertype2016}, the contrast is approximated by
\begin{equation}\label{Eq:SimpleContrast}
    c\negthinspace\p{x, y, v_x, v_y} \propto \sin{\p{\mathcal{A}  \exp\p{-2 \frac{\p{x+v_x t_3}^2 + \p{y+v_y t_3}^2 }{w_0^2} }}}
\end{equation}
$\mathcal{A}$ is the area of the third pulse at the center of the laser beam, which we take as $\pi/2$ in the following. With this expression for the loss of contrast, the calculations of the integrated contrast \eqref{Eq:ContrastLoss} and the contribution of the Gaussian beam curvature \eqref{Eq:GaussianCurvLoss} are presented in \textbf{\nameref{Appendix}}. The analytical series, evaluated up to $n=5$ \cite{series_n5}, are represented by solid lines in Figure~\ref{Fig2:GradiometerFreeFall} and exhibit a behaviour that is in good agreement with the Monte-Carlo simulations. We attribute the difference between this simple analytical model and the exact numerical simulation mainly to the non-uniformity of the coupling during the second pulse. Compared to the results corresponding to the uniform coupling case, displayed as dashed lines, we observe effects of the order of tens of nm~s$^{-2}$ on the gravity bias and of tens of Eötvös on the gravity gradient, which are within reach of current experiments \cite{karcherImprovingAccuracyAtom2018, LyuCompactGraviGradio2022, JanvierCompactDiffGravi2022}.

%%%%%%%%%%%%%%%%%%%%%%%%%%%%%%%%%%%%%%%%%%%%%%%%%%%%%%%%%%%%%%%%%%%%%%%%%%%%%%%%%%%%%%%%%%%%%%
%%%%%%%%%%%%%%%%%%%%%%%%%%  Mirror with Zernike polynomial surface  %%%%%%%%%%%%%%%%%%%%%%%%%%
\section{Mirror with Zernike polynomial surface}\label{SpecificZernike}
In Figure~\ref{Fig2:GradiometerFreeFall}, the rejection of wavefront aberrations of the Gaussian beam is not ideal, even though the Rayleigh length $z_R\approx 100$~m, which is the typical propagation length of the gaussian beam for the considered $w_0=5$~mm, is much larger than the typical longitudinal lengths of the system. We will now study the effect of an additional wavefront aberrations caused by optical surface defects of the retro-reflective mirror. As a simple example, we first assume that the Gaussian beam, the mirror surface, and the two clouds are coaxially centered and that the atomic clouds have no initial transverse velocity. The mirror surface is represented by a Zernike polynomial defined on a reference disk of diameter $2R=27$~mm, of the order of the diameter of our mirror surfaces \cite{pagotInfluenceOpticalAberrations2025}. Only Zernike polynomials with rotation invariance $Z_n^0$, with $n$ even, should be considered, as the others have at least one axis of antisymmetry, which results in a null contribution to the interferometer phase. In this case, an analytical approximation is developed, see \textbf{\nameref{Appendix}}, similar to that presented in \cite{pagotInfluenceOpticalAberrations2025}, which also takes into account the non-uniformity of contrast due to the last pulse, as in the previous case of the Gaussian beam curvature. This analytical approach can be adapted to more complex cases, such as an initially off-centered cloud. With a first-order Taylor expansion of the wavefront aberration and using the series $\mathcal{S}$ defined in expression \eqref{Eq:ZernikeLoss}, the phase bias induced by a mirror with surface $S_\text{mir} = \gamma Z_n^0$ can be approximated by
\begin{equation}\label{Eq:BiasZernikeFiniteWaist}
    \begin{split}
        \delta \phi^{(j)} \approx \p{-1}^\frac{n}{2} k_\text{eff} \, \gamma \Biggl[& \cos{\p{ \frac{\Delta z^{(j)}_{1,\text{mir}}}{l_z^{(n)}} }} \, \mathcal{S}\p{t_1} -2 \cos{\p{ \frac{\Delta z^{(j)}_{2,\text{mir}}}{l_z^{(n)}} }} \, \mathcal{S}\p{t_2} \\
        & + \cos{\p{ \frac{\Delta z^{(j)}_{3,\text{mir}}}{l_z^{(n)}} }} \, \mathcal{S}\p{t_3} \Biggr],
    \end{split}
\end{equation}
with $l_z^{(n)} = 2 k \p{l_{xy}^{(n)}}^2$, $l_{xy}^{(n)} = R/\p{n+1}$ and $\Delta z^{(j)}_{i,\text{mir}}$ the distance between the mirror surface and the position of the atomic cloud during the $i^\text{th}$ pulse. From \eqref{Eq:BiasZernikeFiniteWaist}, the resulting bias on the vertical gravity gradient is
\begin{equation}\label{Eq:vggBias}
    \delta T_{zz} \approx - \p{-1}^\frac{n}{2} \frac{\gamma}{T^2} \sum_{i=1}^3 \xi_i \, \frac{\cos{\p{ \frac{\Delta z^{(l)}_{i,\text{mir}}}{l_z^{(n)}} }} - \cos{\p{ \frac{\Delta z^{(u)}_{i,\text{mir}}}{l_z^{(n)}} }}  }{ z_0^{(l)} - z_0^{(u)} } \, \mathcal{S}\p{t_i},
\end{equation}
with $\xi_1=\xi_3=1$ and $\xi_2=-2$.

\begin{figure}[htbp]
    \centering
    \begin{minipage}{0.49\textwidth}
        \centering
        \includegraphics[trim=0 0.0cm 0 0, clip, width=\linewidth]{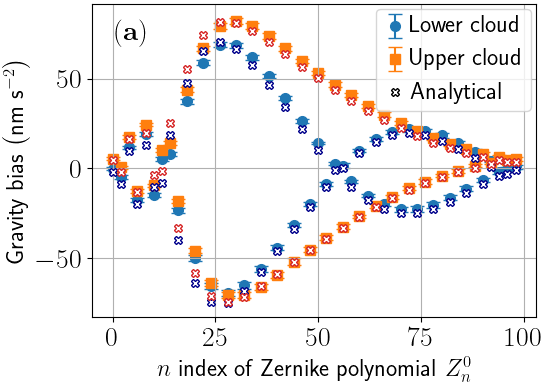}
        %\subcaption{Gravity acceleration bias.}
        %\label{Fig3a:ZernikeRotationInvAcc}
    \end{minipage}%
    \hspace{0.01\textwidth}
    \begin{minipage}{0.49\textwidth}
        \centering
        \includegraphics[trim=0 0.0cm 0 0, clip, width=\linewidth]{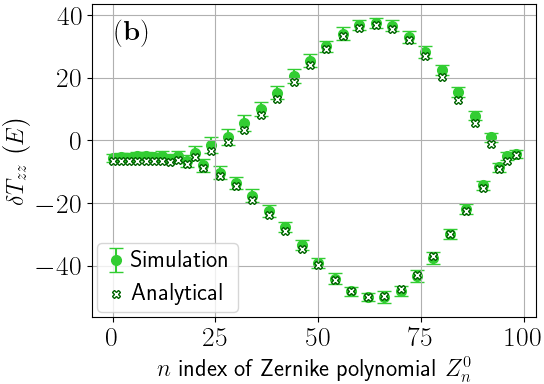}
        %\subcaption{Vertical gravity gradient bias.}
        %\label{Fig3b:ZernikeRotationInvGradient}
    \end{minipage}%
    \caption{Biases in gravity acceleration~(a) and its vertical gradient~(b) for the $2T=400$~ms gradiometer configuration, with a gaussian beam of waist $w_0=5$~mm and a mirror surface proportional to a Zernike polynomial with rotation invariance $S_\text{mir} = \frac{\lambda}{200}Z_n^0$ defined on a disk of diameter $2R = 27$~mm. Each data point is generated from $50$ simulations with $50\thinspace000$ atoms. The crosses correspond to the evaluation of equation \eqref{Eq:BiasZernikeFiniteWaist}, to which the bias of the Gaussian beam calculated with equation \eqref{Eq:GaussianCurvLoss} is added. The series are expanded up to the $5^\text{th}$ order.}
    \label{Fig3:ZernikeRotationInv}
\end{figure}

The results of the numerical simulations for a free fall time $2T=400$~ms are represented as data points in Figure~\ref{Fig3:ZernikeRotationInv}. The calculations of analytical expressions \eqref{Eq:BiasZernikeFiniteWaist} and \eqref{Eq:vggBias} up to $n=5$, in addition to the analytical calculations of the Gaussian beam effect, are represented by crosses. For comparison purposes, the amplitude of the mirror aberration is the same for all polynomials, $\gamma = \lambda/200$, which corresponds to a typical upper limit for a real mirror \cite{pagotInfluenceOpticalAberrations2025}. We note that as long as the amplitude is sufficiently small, \textit{i.e.} $k_\text{eff} \, \gamma \ll 1$~rad, linearization with respect to the amplitude of the aberrations is valid for the full range of Zernike polynomials considered here, since the simulated biases closely correspond to analytical expressions~\eqref{Eq:BiasZernikeFiniteWaist} and \eqref{Eq:vggBias}. The cosines in expression~\eqref{Eq:vggBias} result from the propagation of aberrations, and factorisation with their difference is specific to ideal centered case considered here. When the propagation is neglected or lower-order aberrations are considered, \textit{i.e.}, when the gradiometer baseline is negligible compared to the propagation length: $\left|z_0^{(l)} - z_0^{(u)}\right| \ll l_z^{(n)}$, the rejection can be considered ideal. In this case, the bias on the vertical gravity gradient in Figure~\ref{Fig3:ZernikeRotationInv}~(b) is of $-5.6(1.3)$~E, which corresponds to the contribution of the Gaussian beam reflected on a perfectly flat mirror of $-5.5(5)$~E in Figure~\ref{Fig2:GradiometerFreeFall}. On the other hand, rejection deteriorates when these two lengths are comparable, which, with the parameters used, corresponds to $n \approx 37$, of the order of the threshold observed in the simulations. For higher-order aberrations, the rejection depends on both the cosines difference and the envelope defined by the series $\mathcal{S}$ in expression \eqref{Eq:vggBias}. For $\p{\frac{\sigma_\rho\negthinspace\p{t_2}}{R/(n+1)}}^2 \geq 10$ equivalent to $n\geq 16$, the contribution of wavefront aberrations of the last two pulses of the interferometer is strongly averaged, so that the effect of the first pulse dominates. This is clearly visible in Figure~\ref{Fig3:ZernikeRotationInv}~(a), as the zero of the lower interferometer for $n\approx 54$ equivalent to $l_z^{(n)}=0.97~\text{m} \approx 2 \Delta z^{(l)}_{1,\text{mir}}/\pi$, corresponds to the first cancellation of the cosine in \eqref{Eq:BiasZernikeFiniteWaist}. By replacing the series $\mathcal{S}\p{t_1}$ with $\exp{\p{-\frac{1}{2} \p{\frac{\sigma_\rho\negthinspace\p{t_1}}{l^{(n)}_{xy}}}^2 } }$, its expression in the uniform contrast limit, the maximum bias on the vertical gravity gradient is expected for $n \approx 65$ before decreasing to zero, as in the simulations shown in Figure~\ref{Fig3:ZernikeRotationInv}~(b). 
%
%Although these simulations are performed in the ideal case of centered clouds, we expect that deviations from this ideal case would exhibit similar behaviours for comparable ratio between the typical length scales of the system.

%%%%%%%%%%%%%%%%%%%%%%%%%%%%%%%%%%%%%%%%%%%%%%%%%%%%%%%%%%%%%%%%%%%%%%%%%%%%%%%%%%%%%%%%%%%%%%%%%%%%
%%%%%%%%%%%%%%%%%%%%%%%%%%  Sensitivity to initial transverse conditions  %%%%%%%%%%%%%%%%%%%%%%%%%%
\section{Sensitivity to initial transverse conditions}\label{Sensitivity}
Until now, atomic clouds were considered to be coaxially centered with the Gaussian beam and the mirror surface. In practice, the initial positions of the clouds could be shifted in the transverse plane. They could also have non-zero initial transverse velocities, which would result in variable position offsets over throughout the interferometer. These position offsets must be compared to the typical transverse length of the aberrations under consideration. For instance, in the extreme case where the typical transverse length $l_{xy}$ of the aberrations is much larger than the typical offsets under consideration, the sensitivity to these initial conditions can be neglected.

To evaluate the sensitivity of the experiment to initial transverse conditions, Monte-Carlo simulations are repeated with similar atomic clouds (same widths for position and velocity distribution) whose average transverse conditions $\p{\left\langle \,\vec{r} \, \right\rangle \negthinspace \p{0}, \left\langle \, \vec{v} \, \right\rangle \negthinspace \p{0}}$ are shifted relative to the previous co-centered configuration where we had $\p{\left\langle \, \vec{r} \, \right\rangle \negthinspace \p{0}, \left\langle \, \vec{v} \, \right\rangle \negthinspace \p{0}} = \p{0,0}$. Assuming that there is no preferred direction for the fluctuations in the transverse coordinates, and that the two atomic clouds behave independently, the average initial positions $\p{\left\langle x \right\rangle \negthinspace \p{0}, \left\langle y \right\rangle \negthinspace \p{0}}$ and average initial velocities $\p{\left\langle v_x \right\rangle \negthinspace \p{0}, \left\langle v_y \right\rangle \negthinspace \p{0}}$ of each cloud are drawn from independent centered normal distributions. These distributions are characterized for positions by a standard deviation $\sigma_\rho\negthinspace\p{0} = 0.2$~mm and for velocities by a standard deviation $\sigma_{v_{x,0}, v_{y,0}} = 1$~mm~s$^{-1}$. These fluctuation amplitudes are realistic compared to those measured in experiments \cite{GaugetCharacLimits2009, LuoEvaluatingEffectWA2025}.

\begin{figure}[htbp]
    \centering
    \begin{minipage}{0.49\textwidth}
        \centering
        \includegraphics[trim=0 0.0cm 0 0, clip, width=\linewidth]{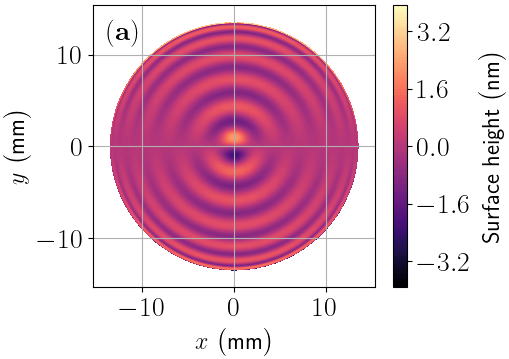}
        %\subcaption{Mirror surface $\frac{\lambda}{200} Z_{25}^{-1}$.}
        %\label{Fig4a:MirrorSurface}
    \end{minipage}%
    \hspace{0.01\textwidth}
    \begin{minipage}{0.49\textwidth}
        \centering
        \includegraphics[trim=0 0.0cm 0 0, clip, width=\linewidth]{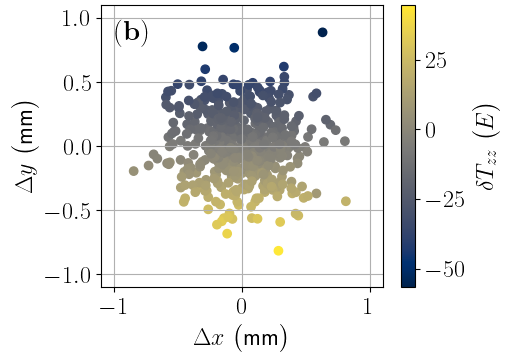}
        %\subcaption{Vertical gravity gradient bias.}
        %\label{Fig4:Sensitivity}~(b)
    \end{minipage}%
    \caption{Simulation of a gradiometer with $2T=400$~ms, $w_0=5$~mm, and with (a) a mirror surface $\frac{\lambda}{200} Z_{25}^{-1}$. (b) $500$~simulations are performed with different initial mean transverse positions chosen from $\mathcal{N}\p{0, \sigma_{x_0, y_0}=0.2~\text{mm}}$ and initial mean transverse velocities chosen from $\mathcal{N}\p{0, \sigma_{v_{x,0}, v_{y,0}} = 1~\text{mm~s}^{-1}}$. Each simulation consists of $50$ evaluations of the interferometers phase with $25\thinspace000$ atoms. The biases are plotted as a function of the difference between the initial mean positions of the two clouds.}
    \label{Fig4:Sensitivity}
\end{figure}

Since the clouds are no longer coaxially centered, Zernike polynomials with anti-symmetric axes will have an effect on measurement bias. An example is shown in Figure~\ref{Fig4:Sensitivity}, where the mirror surface is $S_\text{mir}=\frac{\lambda}{200} Z_{25}^{-1}$. The resulting vertical gravity gradient distribution is characterized by a mean of $-6.2\p{7}$~E and a standard deviation of $16.3\p{5}$~E. The uncertainty on the standard deviation is defined as the squared root of the statistical contribution \cite{uncertainty_variance} and the uncertainty resulting from the fringe fitting. The mean value is consistent with $-5.5(5)$~E, the bias calculated for the interferometer of $2T=400$~ms with the atomic clouds being coaxially centered with the Gaussian beam reflected on a perfectly flat mirror in Figure~\ref{Fig2:GradiometerFreeFall}. Since the main contribution to the bias is imprinted during the first Raman pulse of the interferometer and $\sigma_{v_{x,0}, v_{y,0}} t_1 $ is an order of magnitude smaller than $ \sigma_{x_0, y_0}$, the bias depends mainly on the difference between the initial average positions of the two clouds, as represented in Figure~\ref{Fig4:Sensitivity}~(b). The axis of antisymmetry of the mirror surface is reflected in the sign of the bias of the vertical gravity gradient.

\begin{figure}[h]
    \centering
    \adjincludegraphics[width=\textwidth, trim=0 0 0 1.5cm, clip]{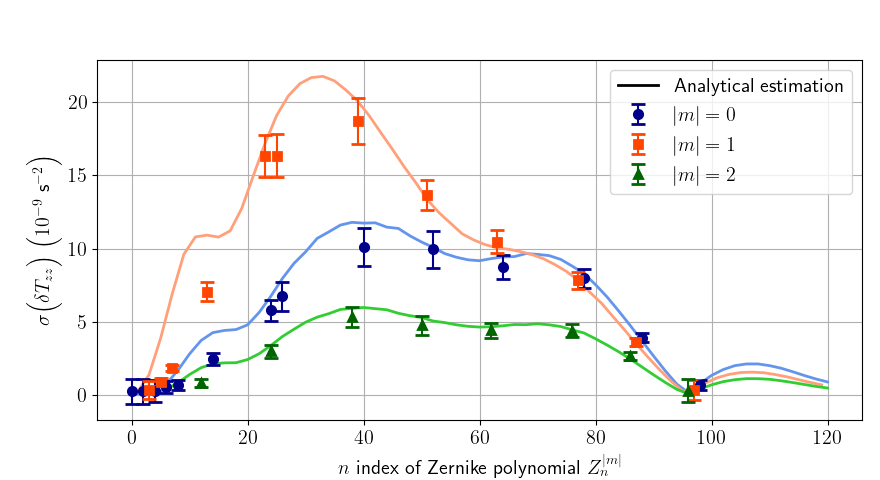}
    \caption{Standard deviation of the vertical gravity gradient $\sigma\p{\delta T_{zz}}$ for a mirror surface $S_\text{mir} = \frac{\lambda}{200} Z_n^m$ and fluctuations in initial transverse positions: $\mathcal{N}\p{0, \sigma_{x_0, y_0}=0.2~\text{mm}}$ and velocities: $\mathcal{N}\p{0, \sigma_{v_{x,0}, v_{y,0}} = 1~\text{mm~s}^{-1}}$. Each data point is calculated over $500$ runs, as in Figure~\ref{Fig4:Sensitivity}~(b), with error bars corresponding to $\pm 3$~times the uncertainty \cite{uncertainty_variance}. Analytical estimates are calculated from $20\thinspace 000$ random initial conditions.}
  \label{Fig5:StandardDev}
\end{figure}

These simulations are repeated for different Zernike polynomials~\cite{Zernike_similar}, and the standard deviations obtained are reported in Figure~\ref{Fig5:StandardDev}. For lower-order aberrations, the impact is lower since the typical lengths of the clouds $\sigma_\rho, \sigma_{x_0, y_0}$ are small compared to the characteristic size $l^{n}_{xy}$ of the aberrations. With these initial transverse fluctuations, the bias on the vertical gravity gradient for Zernike polynomials with rotation invariance $m=0$ and indices $n\leq 20$ can vary by a few Eötvös. This effect is significant compared to the ideal case illustrated in Figure~\ref{Fig3:ZernikeRotationInv}~(b), where only the effect of the gaussian beam is initially visible. 
%Noticeably, the sensitivity increases more rapidly with index $n$ than the biases in Figure~\ref{Fig3:ZernikeRotationInv}~(b). This can be explained by the difference between the initial conditions of the two clouds, such that factorization with the cosines difference in equation~\eqref{Eq:vggBias} is no longer possible, as the associated envelopes are no longer the same. 
For high-order aberrations, with an index $n$ larger than $\approx 100$, the sensitivity returns to zero due to the averaging of aberrations with smaller characteristic transverse lengths $l_{xy}=R/\p{n+1}$.

The dependence of the Zernike polynomials near the center of the disk is
\begin{equation}\label{Eq:ZernikeCenter}
    Z_n^{|m|} \propto \rho^{|m|} \cos{\p{|m|\theta}} + \mathcal{O}\p{\rho^{|m| + 2} \cos{\p{|m|\theta}}}.
\end{equation}
This local behaviour explains why $Z_n^{|m|=1}$, which is locally linear in one direction, is more sensitive to fluctuations in the initial transverse conditions than $Z_n^{m=0}$ and $Z_n^{|m|= 2}$, which are locally quadratic. %On the one hand, near the center, the coefficient of the quadratic term of $Z_n^{|m|= 2}$ is actually larger, by a factor $n/4$, compared to the one of $Z_n^{m=0}$, so that a higher sensitivity could in principle be expected. On the other hand, with a dispersion in the initial transverse position $\sigma_{x_0, y_0}=0.2$~mm smaller than the typical cloud size $\sigma_\rho\negthinspace\p{0} =0.3$~mm, the effect of $Z_n^{|m|= 2}$ is actually reduced, because the cloud distribution generally crosses an axis of antisymmetry. For Zernike polynomials with higher $m$ indexes, locally flatter and featuring more axes of antisymmetry, the biases caused by wavefront aberrations will thus be even less pronounced. 
Moreover, near the center, the coefficient of the quadratic term of $Z_n^{|m|= 2}$ is actually smaller, by a factor of $1/2$, than that of $Z_n^{m=0}$, which corresponds approximately to the ratio of the calculations in Figure~\ref{Fig5:StandardDev}. For Zernike polynomials with higher $m$ indices, which are locally flatter, the biases caused by wavefront aberrations will thus be even less pronounced. This definition of the sensitivity of the measurement to mirror surface aberrations described by Zernike polynomials in the linear amplitude regime could be used to define the mirror specifications.

To calculate analytical estimates of the standard deviation for clouds that initially have non-zero transverse conditions, analytical formulas \eqref{Eq:BiasZernikeFiniteWaist} and \eqref{Eq:ZernikeLoss} must be extended. As indicated in \cite{pagotInfluenceOpticalAberrations2025}, within the limit of uniform contrast, for a given Zernike polynomial $Z_n^m$ and for an atomic cloud close to the center of the mirror surface, \textit{i.e.} $\sqrt{ \langle x \rangle^2 + \langle y \rangle^2} \ll R$, the exponential term describing the averaging of aberrations is identical to that of a centered cloud and for the Zernike polynomial $Z_n^0$, so that the bias is simply multiplied by the value of the Zernike polynomial defined at the average position of the cloud $\p{\langle x \rangle, \langle y \rangle}$. Assuming that this relation remains valid in the present case, since the displacements considered are small compared to the mirror radius and the beam waist: $\sigma_{x_0, y_0}, \sigma_{v_{x,0}, v_{y,0}} t_3 \ll R, w_0$, the formula for the bias \eqref{Eq:BiasZernikeFiniteWaist} is extended to
\begin{equation}\label{Eq:BiasZernikeFiniteWaist_Generalized}
    \begin{split}
        \delta \phi^{(j)}\left[ Z_n^m \right] = k_\text{eff} \, \gamma \Biggl[\phantom{-2}& Z_n^m \negthinspace \p{\left\langle x\p{t_1} \right \rangle, \left\langle y\p{t_1} \right \rangle} \, \cos{\p{ \frac{\Delta z^{(j)}_{1,\text{mir}}}{l_z^{(n)}} }} \, \mathcal{S}\p{t_1} \\
        -2 &Z_n^m \negthinspace \p{\left\langle x\p{t_2} \right \rangle, \left\langle y\p{t_2} \right \rangle} \, \cos{\p{ \frac{\Delta z^{(j)}_{2,\text{mir}}}{l_z^{(n)}} }} \, \mathcal{S}\p{t_2} \\
        + &Z_n^m \negthinspace \p{\left\langle x\p{t_3} \right \rangle, \left\langle y\p{t_3} \right \rangle} \, \cos{\p{ \frac{\Delta z^{(j)}_{3,\text{mir}}}{l_z^{(n)}} }} \, \mathcal{S}\p{t_3} \Biggr].
    \end{split}
\end{equation}
$\p{\left\langle x\p{t_i} \right \rangle, \left\langle y\p{t_i} \right \rangle}$ are the average transverse positions of the cloud at the time of the $i^\text{th}$ pulse. The standard deviation of the vertical gravity gradient values estimated using this formula is displayed as solid lines in Figure~\ref{Fig5:StandardDev} on $20\thinspace 000$ random initial conditions. The results agree well with the numerical simulations. The largest deviations are observed for low $n$ indices, which can be explained by the replacement of Zernike radial polynomials by Bessel functions, which is more accurate for high $n$ indices.

%%%%%%%%%%%%%%%%%%%%%%%%%%%%%%%%%%%%%%%%%%%%%%%%%%%%%%%%%%%%%%%
%%%%%%%%%%%%%%%%%%%%%%%%%%  Scaling  %%%%%%%%%%%%%%%%%%%%%%%%%%
\section{Scaling}\label{Scaling}
In this section, we discuss how biases on the measured gravity acceleration and its vertical gradient vary as a function of experimental parameters.
%
%In general, since the Zernike polynomials and cosines in the bias expression \eqref{Eq:BiasZernikeFiniteWaist_Generalized} are bounded by $1$ (in absolute value), the total contribution is less than the sum of the series defined in expression~\eqref{Eq:ZernikeLoss} multiplied by the coefficient $k_\text{eff} \, \gamma$. Even in this worst-case scenario, the bias on the gravity measurement and on the vertical gravity gradient measurement are both inversely proportional to the square of the free fall time $2T$ of the interferometer, and the bias on the gravity gradient is inversely proportional to the baseline $z_0^{(l)} - z_0^{(u)}$ of the gradiometer. Moreover, the averaging of aberrations over the cloud size remains present.

First, scaling with the interferometer geometry parameters, the free-fall time and the baseline of the gradiometer, can be deduced either from expression~\eqref{Eq:PhaseShiftGen}, by majoring the phase of each atom by the peak-to-valley of the phase fluctuations or from equation~\eqref{Eq:BiasZernikeFiniteWaist_Generalized} as Zernike polynomials and cosines are bounded by $1$ in absolute value. Denoting $k_\text{eff} \Delta \gamma$ the maximum amplitude of the phase fluctuations at the three pulses, the biases in gravity accelerations and vertical gradient are bounded by
\begin{equation}\label{Eq:Bias_max}
    \left| \delta g^{(j)} \right| \leq \frac{4 \Delta \gamma}{T^2}  \text{ and } \left| \delta T_{zz} \right| \leq \frac{4 \Delta \gamma}{T^2 \p{z_0^{(u)} - z_0^{(l)}} }.
\end{equation}
The factor $4$ results from the sum of the laser phases at the three pulses, the phase of the laser at the mirror pulse being imprinted twice. Thus, increasing the free-fall time and the baseline of the gradiometer allows to reduce the biases in this worst case scenario. Though, even with the chosen parameters, $2T=400$~ms and $z_0^{(u)} - z_0^{(l)}=1$~m necessitating a $2$-meter-high experiment, expression~\eqref{Eq:Bias_max} with $\Delta \gamma=\lambda/200$ corresponds to a bias on the vertical gravity gradient of $390$~E. Hopefully, experimentally the bias is lower as correlation between the phases fluctuations with large transverse aberrations length and efficient averaging of those with typical transverse length smaller than the atomic cloud size allows for a decrease of the wavefront aberrations effect as shown in Figure~\ref{Fig3:ZernikeRotationInv}.

\begin{figure}[ht!]
     \centering
     \begin{subfigure}[b]{0.42\textwidth}
         \centering
         \includegraphics[height=5.1cm]{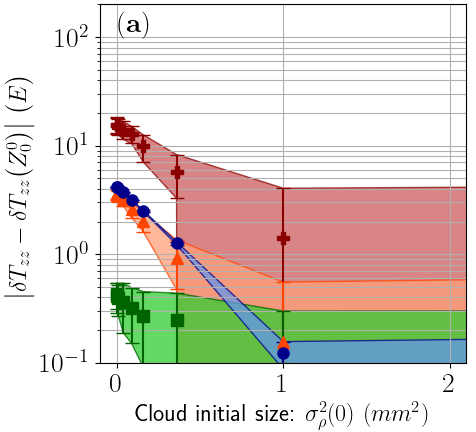}
         %\caption{Mirror surface $S_\text{mir} = \frac{\lambda}{200} Z_{34}^0$}
         %\label{Fig6:ScalingSizeTemp}~(a)
     \end{subfigure}
     \begin{subfigure}[b]{0.33\textwidth}
         \centering
         \includegraphics[height=5.1cm, trim=2.3cm 0cm 0cm 0cm, clip=true]{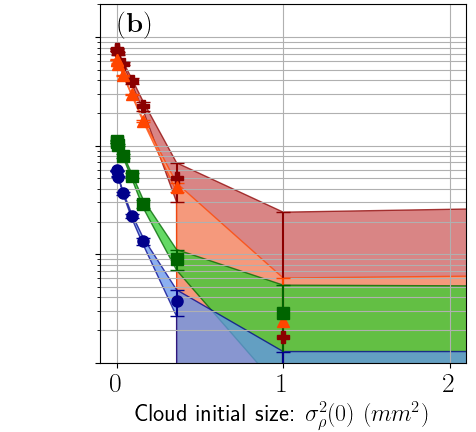}
         %\caption{Mirror surface $S_\text{mir} = \frac{\lambda}{200} Z_{54}^0$}
         %\label{Fig6:ScalingSizeTemp}~(b)
     \end{subfigure}
     \begin{subfigure}[b]{0.02\textwidth}
         \centering
         \raisebox{2.5cm}{\includegraphics[width=2cm]{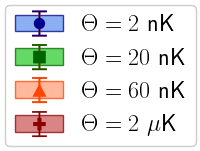}}
     \end{subfigure}
     \hspace{1cm}
     \caption{Absolute value of the vertical gravity gradient bias for a mirror surface (a)~$S_\text{mir}= \frac{\lambda}{200} Z_{34}^0$ and (b)~$S_\text{mir} = \frac{\lambda}{200} Z_{54}^0$ without fluctuations in initial conditions (positions and velocities) as a function of the square of the initial cloud size $\sigma_\rho\negthinspace\p{0}$ and for different temperatures $\Theta \in \left\{ 2,\ 20,\ 60,\ 2000  \right\}$~nK. The shaded areas represent uncertainties. Each value is determined from $50$ simulations of the experiment with $50 \thinspace 000$ atoms in each cloud.}
     \label{Fig6:ScalingSizeTemp}
\end{figure}

To study the scaling of the wavefront aberrations effect with the kinematic parameters of the cloud, we keep the geometry of the gradiometer defined above and we chose to examine the vertical gravity gradient bias for mirror surfaces described by Zernike polynomials $Z_{34}^{0}$ and $Z_{54}^0$ with the same amplitude $\lambda/200$, which cause biases of the order of tens of Eötvös, as shown in Figure~\ref{Fig3:ZernikeRotationInv}~(b). For atomic clouds axially co-centered with the mirror surface and the laser beam, and with no average transverse velocity, the bias on the gravity gradient is calculated as a function of the initial size $\sigma_\rho\negthinspace\p{0}$ of the cloud for different temperatures in the range $\left[2\text{ nK},\thinspace 2~\mu\text{K} \right]$. In order to focus on the contribution of the mirror surface, the contribution of the Gaussian beam is eliminated by considering the difference with simulations performed with a perfectly flat mirror $S^{\,\text{flat}}_\text{mir} \propto Z_0^0$. The absolute value of the difference is represented in solid symbols in Figure~\ref{Fig6:ScalingSizeTemp}. Finally, for clarity, for each temperature, the error bars defining the uncertainties in the simulation results have been connected and the resulting areas filled with the corresponding color, as shown in Figure~\ref{Fig6:ScalingSizeTemp}. These uncertainties represent the standard deviation of the results of $50$ experiments with $50\thinspace000$ atoms in each clouds.

Regardless of temperature, all simulations in Figure~\ref{Fig6:ScalingSizeTemp} show an exponential decrease proportional to the square of the initial size $\sigma_\rho\negthinspace\p{0}$ of the clouds, and the slope in Figure~\ref{Fig6:ScalingSizeTemp}~(b) for $Z_{54}^0$ is higher than in Figure~\ref{Fig6:ScalingSizeTemp}~(a) for $Z_{34}^0$. This agrees with the analytical expression~\eqref{Eq:ZernikeLoss}, as this decrease is faster for aberrations with a smaller typical transverse size (with $l_{xy} \propto 1/\p{n+1}$), due to averaging over the size of the cloud. Thus, when the clouds initial size $\sigma_\rho \negthinspace \p{0}$ is large enough so that the effects of the mirrors surface are averaged to zero, only the effect due to the curvature of the gaussian beam remains and, as it has been subtracted in Figure~\ref{Fig6:ScalingSizeTemp}, the difference is compatible with zero within the error bars.

The temperature dependence is more complex. While the temperature-dependent curves are ordered in Figure~\ref{Fig6:ScalingSizeTemp}~(b) for the mirror surface proportional to $Z^0_{54}$, this is not the case in Figure~\ref{Fig6:ScalingSizeTemp}~(a) for the mirror surface proportional to $Z^0_{34}$. At lower temperatures, the cloud expands less during its free fall, so that the averaging during the second and third pulses is similar to that of the first pulse. Depending on the cosine terms related to propagation in equation~\eqref{Eq:vggBias}, these contributions can either increase or decrease the total bias. As illustrated in Figure~\ref{Fig6:ScalingSizeTemp}, the uncertainties represented by the shaded areas increase as the clouds get larger and experience more inhomogeneities from the laser beam.

%%%%%%%%%%%%%%%%%%%%%%%%%%%%%%%%%%%%%%%%%%%%%%%%%%%%%%%%%%%%%%%%%%%%%%%%%%%%%%%%%%%%%%%%%%%%
%%%%%%%%%%%%%%%%%%%%%%%%%%  Simulations with real mirror surface  %%%%%%%%%%%%%%%%%%%%%%%%%%
\section{Simulations with real mirror surface}\label{RealSurface}
\begin{figure}[h]
    \centering
    \begin{minipage}{0.42\textwidth}
        \centering
        \begin{minipage}{\textwidth}
            \includegraphics[width=\linewidth, trim=0.6cm 0.1cm 0.5cm 4.2cm, clip=true]{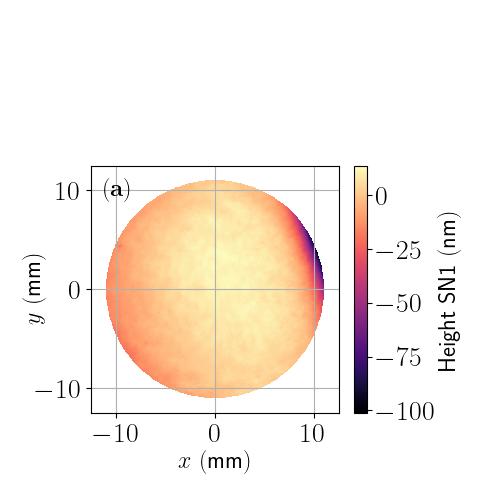}
            %\subcaption{Mirror surface SN1}
            %\label{Fig7:GradioRealMir}~(a)
        \end{minipage}
        \vspace{0.5em} % Ajuste l'espace entre les images
        \begin{minipage}{\textwidth}
            \includegraphics[width=\linewidth, trim=0.6cm 0.1cm 0.5cm 3.8cm, clip=true]{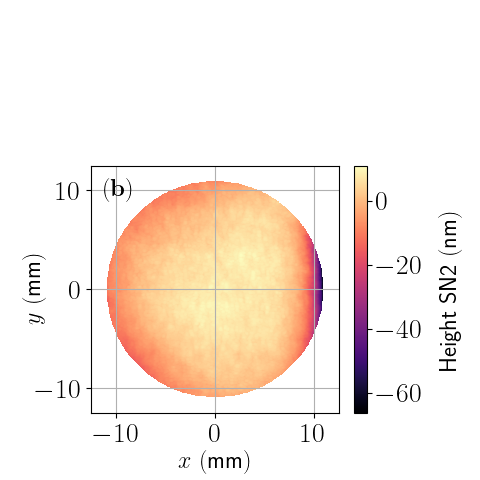}
            %\subcaption{Mirror surface SN2}
            %\label{Fig7:GradioRealMir}~(b)
        \end{minipage}
        \vspace{0.5em} % Ajuste l'espace entre les images
        \begin{minipage}{\textwidth}
            \includegraphics[width=\linewidth, trim=0.6cm 0.1cm 0.5cm 4.2cm, clip=true]{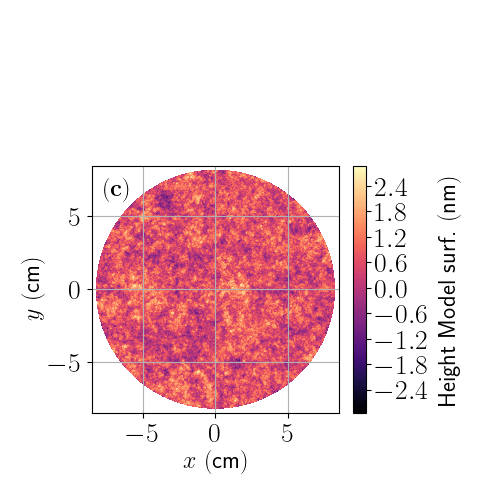}
            %\subcaption{Mirror PSD model}
            %\label{Fig7:GradioRealMir}~(c)
        \end{minipage}
    \end{minipage}
    \hspace{0.01\textwidth}
    \begin{minipage}{0.555\textwidth}
        \centering
        \includegraphics[width=\linewidth, trim=0cm 0cm 0.2cm 1.0cm, clip=true]{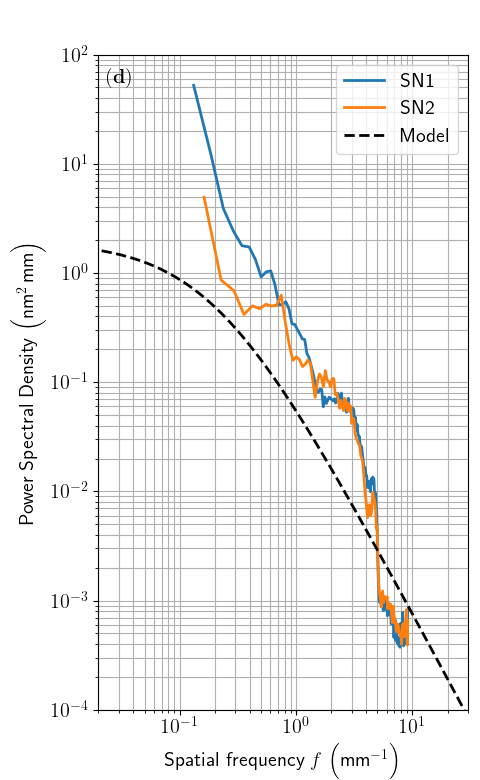}
        %\subcaption{Mirror surface PSD}
        %\label{Fig7:GradioRealMir}~(d)
    \end{minipage}%
    \caption{Map of the real mirror surfaces SN1 (a) and SN2 (b) measured using a Fizeau interferometer, with their PSD (d). (c) Example of a mirror surface generated numerically using the PSD model, similar to \cite{HiroseSapphireMirror2014}, represented by the black dashed line in (d).}
    \label{Fig7:GradioRealMir}
\end{figure}

So far, we have examined a simplified model based on a mirror surface described by a single Zernike polynomial with a fixed amplitude. In this section, the contribution to the gravitational acceleration and gradient bias is calculated from an actual mirror surface measured using a Fizeau interferometer \cite{FormanZygoInterferometer1979}. The surfaces shown in Figures~\ref{Fig7:GradioRealMir}~(a) and~\ref{Fig7:GradioRealMir}~(b) are the same as in \cite{pagotInfluenceOpticalAberrations2025}, and their power spectral density~\cite{PSD_1D} (PSD) is illustrated in Figure~\ref{Fig7:GradioRealMir}~(d).

As in section~\textbf{\nameref{Sensitivity}}, $500$~numerical simulations are used to evaluate the sensitivity of the measurements to typical variations in the initial transverse positions $\sigma_{x_0, y_0}=0.2~\text{mm}$ and transverse velocities $\sigma_{v_{x,0}, v_{y,0}} = 1~\text{mm~s}^{-1}$. In addition, since the mirror surfaces have been decomposed into Zernike polynomials \cite{pagotInfluenceOpticalAberrations2025}, estimates of the mean and standard deviation of the bias distributions~\cite{Mirror_decomposition} can also be calculated from the Zernike coefficients using formula~\eqref{Eq:BiasZernikeFiniteWaist_Generalized}. 
\begin{table}
    \caption{Mean and standard deviation of the bias on the gravity acceleration and its vertical gradient with mirror surfaces SN$1$ and SN$2$.$^{a,b}$}
    \label{Tab1:GradioRealMir}
    \centering
    \begin{tabular}{|r|r|r|r|r|}
        \hline
         & \adjustbox{valign=c, padding=0.25ex 1ex 0.25ex 1ex}{SN1 num.} & \adjustbox{valign=c, padding=0.25ex 1ex 0.25ex 1ex}{SN1 Zer.} & \adjustbox{valign=c, padding=0.25ex 1ex 0.25ex 1ex}{SN2 num.} & \adjustbox{valign=c, padding=0.25ex 1ex 0.25ex 1ex}{SN2 Zer.} \\
         \hline
        \adjustbox{valign=c, padding=0.25ex 1ex 0.25ex 1ex}{$\left\langle \delta g^{(l)} \right\rangle\ \p{\text{nm s}^{-2}}$}           & \adjustbox{valign=c, padding=1ex 1ex 1ex 1ex}{$7.6(1)$}  & \adjustbox{valign=c, padding=1ex 1ex 1ex 1ex}{$10.7(7)$}   & \adjustbox{valign=c, padding=1ex 1ex 1ex 1ex}{$6.1(2)$} & \adjustbox{valign=c, padding=1ex 1ex 1ex 1ex}{$5.6(2)$} \\
        \hline
        \adjustbox{valign=c, padding=0.25ex 1ex 0.25ex 1ex}{$\sigma\p{\delta g^{(l)}}\ \p{ \text{nm s}^{-2}}$} & \adjustbox{valign=c, padding=1ex 1ex 1ex 1ex}{$1.7(1)$}  & \adjustbox{valign=c, padding=1ex 1ex 1ex 1ex}{$3.3(7)$}    & \adjustbox{valign=c, padding=1ex 1ex 1ex 1ex}{$3.2(1)$} & \adjustbox{valign=c, padding=1ex 1ex 1ex 1ex}{$5.5(2)$} \\
        \hline
        \adjustbox{valign=c, padding=0.25ex 1ex 0.25ex 1ex}{$\left\langle \delta g^{(u)} \right\rangle\ \p{ \text{nm s}^{-2}}$}           & \adjustbox{valign=c, padding=1ex 1ex 1ex 1ex}{$18.1(2)$}  & \adjustbox{valign=c, padding=1ex 1ex 1ex 1ex}{$22.2(7)$}  & \adjustbox{valign=c, padding=1ex 1ex 1ex 1ex}{$8.7(2)$} & \adjustbox{valign=c, padding=1ex 1ex 1ex 1ex}{$9.8(2)$} \\
        \hline
        \adjustbox{valign=c, padding=0.25ex 1ex 0.25ex 1ex}{$\sigma\p{\delta g^{(u)}}\ \p{ \text{nm s}^{-2}}$} & \adjustbox{valign=c, padding=1ex 1ex 1ex 1ex}{$5.1(1)$}   & \adjustbox{valign=c, padding=1ex 1ex 1ex 1ex}{$6.4(7)$}   & \adjustbox{valign=c, padding=1ex 1ex 1ex 1ex}{$3.7(1)$} & \adjustbox{valign=c, padding=1ex 1ex 1ex 1ex}{$5.6(2)$} \\
        \hline
        \adjustbox{valign=c, padding=0.25ex 1ex 3.25ex 1ex}{$\left\langle \delta T_{zz} \right\rangle\ \p{E}$}              & \adjustbox{valign=c, padding=1ex 1ex 1ex 1ex}{$-10.5(3)$} & \adjustbox{valign=c, padding=1ex 1ex 1ex 1ex}{$-11.4(9)$} & \adjustbox{valign=c, padding=1ex 1ex 1ex 1ex}{$-2.7(2)$} & \adjustbox{valign=c, padding=1ex 1ex 1ex 1ex}{$-4.2(3)$} \\
        \hline
        \adjustbox{valign=c, padding=0.25ex 1ex 3.25ex 1ex}{$\sigma\p{\delta T_{zz}}\ \p{E}$}    & \adjustbox{valign=c, padding=1ex 1ex 1ex 1ex}{$5.4(1)$}   & \adjustbox{valign=c, padding=1ex 1ex 1ex 1ex}{$7.2(9)$}   & \adjustbox{valign=c, padding=1ex 1ex 1ex 1ex}{$5.0(2)$} & \adjustbox{valign=c, padding=1ex 1ex 1ex 1ex}{$7.8(3)$} \\
        \hline
    \end{tabular}\\
    $^a$The results of numerical simulations (\textit{num.}) are calculated over $500$~runs. The results obtained with Zernike decomposition of the surfaces (for polynomials with $|m|\in\left\{ 0, 1, 2\right\}$ up to order $2\thinspace500$ in ISO-$14999$) and the extrapolated equation \eqref{Eq:BiasZernikeFiniteWaist_Generalized} (\textit{Zer.}) are calculated over $20\thinspace000$~runs.\\
    $^b$Uncertainties in (\textit{Zer.}) result from the calculation of the coefficients $\p{c_i \pm \delta c_i}_{1\leq i \leq 2\thinspace500}$ of the Zernike polynomials. As changing the coefficients leads to correlated errors for every atoms, the uncertainty is calculated by performing the quadratic sum of \eqref{Eq:BiasZernikeFiniteWaist_Generalized} with amplitudes $\p{\delta c_i}_{1\leq i \leq 2\thinspace500}$.
\end{table}
The results of the simulations and analytical estimates with the surfaces of the two mirrors are summarized in Table~\ref{Tab1:GradioRealMir}. The average biases obtained from the numerical and analytical calculations are marginally consistent, given their uncertainties. As for the standard deviations, those estimated analytically are generally larger than those simulated, which we attribute to the overestimation in the formula~\eqref{Eq:BiasZernikeFiniteWaist_Generalized} of the effect of lower-order Zernike polynomials, as illustrated in Figure~\ref{Fig5:StandardDev}. Nevertheless, when the Zernike decomposition of the mirror surface is available, equation~\eqref{Eq:BiasZernikeFiniteWaist_Generalized} can be used to quickly estimate the bias induced by surface defects as well as an upper limit of the expected standard deviation for given initial transverse distributions. The average bias and associated standard deviation of the order of ten Eötvös on gravity gradient measurements with mirror surfaces~\ref{Fig7:GradioRealMir}~(a) and \ref{Fig7:GradioRealMir}~(b), correspond to the typical errors encountered in existing experiments~\cite{LyuCompactGraviGradio2022}.

%%%%%%%%%%%%%%%%%%%%%%%%%%%%%%%%%%%%%%%%%%%%%%%%%%%%%%%%%%%%%%%%%%%%%%%%%%%%%%%%%%%%%%%%%%
%%%%%%%%%%%%%%%%%%%%%%%%%%  Simulations for a 10 m gradiometer  %%%%%%%%%%%%%%%%%%%%%%%%%%
\section{Simulations for a $\mathbf{10}$ m gradiometer}\label{10mGradio}
To conclude this study on wavefront aberrations, we perform simulations in a state-of-the-art $10$-meter-long gradiometer \cite{HartwigTestingUFF2015, AsenbaumPhaseShiftSpacetimeCurvature2017}. The beam has a radius at $1/e^2$ in intensity of $w_0=2.3$~cm, while the initial transverse size is $\sigma_\rho\negthinspace\p{0}=300~\mu$m and the expansion temperature is $50$~nK for both clouds. The mirror is placed at a height $z=10.10$~m, its diameter is $2R=16.4$~cm and its surface, assumed to be isotropic, is generated according to the PSD theoretical model in Figure~\ref{Fig7:GradioRealMir}~(d). This theoretical model, similar to that of~\cite{HiroseSapphireMirror2014}, is significantly better than the real mirror surfaces SN1 and SN2 shown in Figures~\ref{Fig7:GradioRealMir}~(a) and \ref{Fig7:GradioRealMir}~(b), although it corresponds to existing mirrors \cite{HiroseSapphireMirror2014, DroriScatteringLossPrecisionMetrology2022}. As in previous simulations, the two atomic clouds are prepared at two positions separated by the baseline length. They are then released, with zero initial average longitudinal velocity, and fall into the gravitational field. At the end of the interferometer sequence, which lasts $2T=1$~s and starts at $t_1=15.8$~ms, they have traveled an approximate distance of $5$~m. In the following simulations, where the distance between the two clouds will be modified, the upper cloud is always assumed to be prepared at a height $z=10$~m.

\begin{figure}[h]
    \centering
    \adjincludegraphics[width=\textwidth, trim=0 0 0 1.5cm, clip]{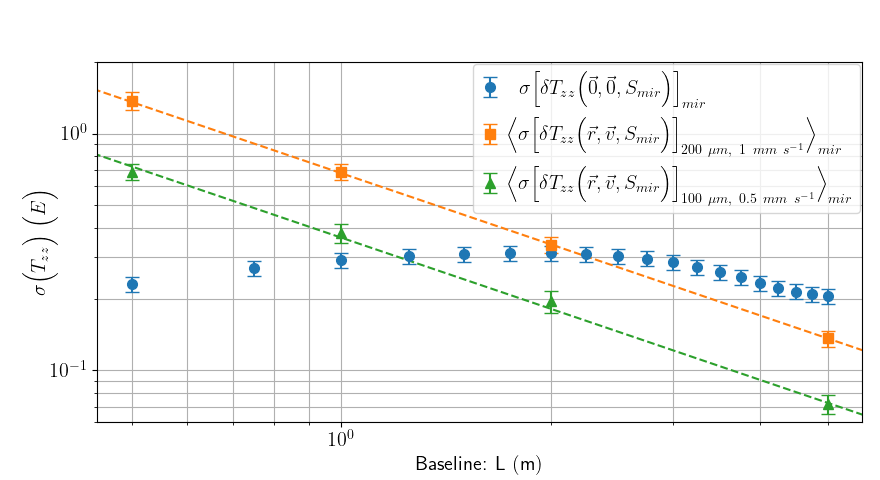}
    \caption{Standard deviation of the vertical gravity gradient for different baseline lengths. (blue dots) The standard deviation is calculated over $100$ mirror surfaces for coaxially centered clouds. Standard deviation obtained for clouds with initial transverse fluctuations (orange squares) $\left\{\sigma_{x_0, y_0}=0.2~\text{mm}, \ \sigma_{v_{x,0}, v_{y,0}} = 1~\text{mm~s}^{-1} \right\}$ and (green triangles) $\left\{\sigma_{x_0, y_0}=0.1~\text{mm},\ \sigma_{v_{x,0}, v_{y,0}} = 0.5~\text{mm~s}^{-1} \right\}$ and averaged over $20$ mirror surfaces.}
  \label{Fig8:Std}
\end{figure}

First, in the ideal case where both atomic clouds are co-centered with the mirror and the laser beams, \textit{i.e.} $\left\langle \, \Vec{r} \, \right\rangle \p{0}=\Vec{0}$, and have no average transverse velocity, $\left\langle \, \Vec{v} \, \right\rangle \p{0}=\Vec{0}$, the bias on the vertical gravity gradient is calculated for different baseline lengths in the range $\left[0.5,\thinspace 5 \right]$~m. The typical magnitude of the bias for a random mirror surface with the same PSD is evaluated by calculating the standard deviation over $100$ mirror surfaces. It is displayed as blue dots in Figure~\ref{Fig8:Std}. Remarkably, in the range of baselines explored here, its scaling is not proportional to the inverse of the baseline length. This can be explained by the bias expression~\eqref{Eq:vggBias}, since the difference between the cosine terms that account for the propagation of the aberrations also depends on the baseline. For the PSD model considered here, even in this ideal case, the bias on the gravity gradient is typically of the order of $0.1$~E, which corresponds to the order of magnitude of the accuracy required for the most demanding scenario in~\cite{HartwigTestingUFF2015}.

On top of the bias in the ideal case due to the specific defects of the mirror surface, additional contributions arise from fluctuations in the initial transverse conditions of the atomic clouds. As before, typical fluctuations in position $\left\langle \, \Vec{r} \, \right\rangle \p{0} \sim \mathcal{N}\p{0, \sigma_{x_0, y_0}=0.2~\text{mm}}$ and in velocity $\left\langle \, \Vec{v} \, \right\rangle \p{0} \sim \mathcal{N}\p{0, \sigma_{v_{x,0}, v_{y,0}} = 1~\text{mm~s}^{-1}}$ are assumed. Since typical defect fluctuations are similar for mirror surfaces with the same PSD, the phase bias standard deviation caused by fluctuations of initial conditions should also be similar for different mirror surfaces of this type. We therefore calculate the standard deviations over $100$ runs with fluctuating positions $\left\langle \, \Vec{r} \, \right\rangle \p{0}$ and velocities $\left\langle \, \Vec{v} \, \right\rangle \p{0}$ for $20$ different synthetic mirror surfaces. These $20$ values are eventually averaged and represented by the orange squares in Figure~\ref{Fig8:Std}. The contributions of the ideal case of a gaussian beam reflected on a perfectly flat mirror are not subtracted, as they are two orders of magnitude smaller. Notably, the standard deviation related to these initial transverse conditions fluctuations decreases linearly with the baseline length $L$. Adjusting these points to the function $\sigma_\text{eff}\left[\sigma_{x_0, y_0}, \sigma_{v_{x,0}, v_{y,0}}\right]/\p{L T^2}$, represented by the orange dashed line in Figure~\ref{Fig8:Std}, gives an effective height fluctuation $\sigma_\text{eff}\left[200~\mu\text{m}, 1\text{ mm s}^{-1}\right] = 0.1700(3)$~nm. Similarly, simulations with fluctuations in position $\sigma_{x_0, y_0}=0.1~\text{mm}$ and in velocity $\sigma_{v_{x,0}, v_{y,0}} = 0.5~\text{mm~s}^{-1}$ were performed, although in this case, $20$ different mirror surfaces were used for each baseline length. The corresponding results, represented by green triangles in Figure~\ref{Fig8:Std}, still show behaviour inversely proportional to the baseline length, and the fit gives an effective height fluctuation $\sigma_\text{eff}\left[100~\mu\text{m}, 0.5\text{ mm s}^{-1}\right] = 0.091(2)$~nm. This decrease in the mean standard deviation is mainly due to the decrease in the initial velocity distribution, as simulations with the parameters $\sigma_{x_0, y_0}=0.1~\text{mm}$ and $\sigma_{v_{x,0}, v_{y,0}} = 1~\text{mm~s}^{-1}$, not shown in Figure~\ref{Fig8:Std}, give an effective height fluctuation $\sigma_\text{eff}\left[100~\mu\text{m}, 1\text{ mm s}^{-1}\right] = 0.161(3)$~nm.

Thus, in this $10$-meter-long configuration and with a mirror whose PSD is defined in Figure~\ref{Fig7:GradioRealMir}~(d), it is possible to achieve a gravity gradient accuracy of a few $0.1$~E, which corresponds to the most stringent requirement in \cite{HartwigTestingUFF2015}, given that the fluctuations in the initial transverse conditions are less than the realistic values considered here \cite{GaugetCharacLimits2009, LuoEvaluatingEffectWA2025}.

%%%%%%%%%%%%%%%%%%%%%%%%%%%%%%%%%%%%%%%%%%%%%%%%%%%%%%%%%%%%%%%%%%
%%%%%%%%%%%%%%%%%%%%%%%%%%  Conclusion  %%%%%%%%%%%%%%%%%%%%%%%%%%
\section{Conclusion}\label{Conclusion}
We have demonstrated that the rejection of wavefront aberrations in an atomic gradiometer is finite, and that the residual bias on the measurement of the vertical gravity gradient is within the reach of current experiments. Besides, in an interferometric configuration where an atomic cloud expands up to the size of the laser beam, contrast non-uniformity plays a crucial role in the phase bias of the interferometer. This applies to both the gaussian beam curvature and the aberrations caused by mirror surface defects. Within this limitation, the analytical framework developed in~\cite{pagotInfluenceOpticalAberrations2025} has been extended to allow rapid estimation of the induced biases and can be adapted to atomic interferometers of different geometries. In addition, simulations with fluctuations in the initial transverse conditions of the atomic clouds allowed us to estimate the sensitivity of the measurement to the initial kinematic parameters of the atomic sources. Eventually, in a state-of-the-art configuration, with a high quality mirror such as that designed for optical gravitational wave detectors \cite{HiroseSapphireMirror2014, DroriScatteringLossPrecisionMetrology2022}, we show that wavefront aberrations could cause a bias of the order of $0.1$~E on the measurement of the vertical gravity gradient, which corresponds to the accuracy requirement targeted in some experiments~\cite{HartwigTestingUFF2015}.

Since we focused on the effects of wavefront aberrations, we did not take into account the Coriolis effect, which also contributes to the interferometer phase \cite{LanInfluenceOfCoriolis2012, hogan2008lightpulseatominterferometry,LyuCompactGraviGradio2022}, particularly when considering fluctuations in the initial transverse velocity. Moreover, for simplicity, the impact of detection on the interferometer phase was omitted. In particular, in the expression~\eqref{Eq:PhaseIdeal}, the term proportional to $\propto T_{zz}^2 T_\text{det}$, where $T_\text{det}$ is the delay between the last pulse and detection, has been neglected~\cite{APetersHighPrecision2001}. More importantly, the finite size of the detection area may also play a role, when certain classes of atoms are not detected or have lower weights~\cite{gillotLimitsSymmetryMachZehndertype2016, karcherImprovingAccuracyAtom2018}. Furthermore, our simulations are based on calculating laser beam distortions at the average longitudinal position of the atoms at each pulse. Since the typical separation distance along the optical axis is $\Delta z =v_\text{rec} T$, aberrations with a comparable typical propagation length correspond to a typical transverse length $\sqrt{\Delta z / k_\text{eff}}$, which corresponds to $20~\mu$m in the case of the $10$-meter-long gradiometer. This value is smaller than the size of the pixels used to represent the mirror surfaces and is an order of magnitude smaller than the transverse size of the atomic clouds considered, so their effect, once averaged, should be negligible. However, for initially smaller clouds, \textit{e.g.} those prepared in a dipolar trap, the corrections could be significant. Moreover, although Zernike polynomials offer the advantage of analytical approximations and enable us to understand the role of key experimental parameters, characterizing optics surfaces at length scales well below the radius on which these polynomials are defined is a tedious task. An alternative is to use the PSD to characterize the mirror surface. In this case, and within the limits of small defects, an analytical approach based on Sinusoidal-Gaussian beams~\cite{Casperson97} for the propagation of aberrations might be developed. This would be particularly useful for defining specifications of the optics used in experiments, especially since the PSD can be reconstructed from different measurements at different length scales \cite{JacobsQuatitativeCharacSurface2017, HiroseSapphireMirror2014, DroriScatteringLossPrecisionMetrology2022}. 

Although the primary objective of this work is to improve the characterization of our gradiometry experiment~\cite{CaldaniSimultaneousGraviGradio2019}, it can be adapted to other experiments with different geometries, such as a four-pulse cold atom gyroscope~\cite{GautierAccurateMeasurementSagnac2022}. In this case, even if the atomic cloud is ideally located at the same mean position in the laser beams during the acending and descending pulses, and if wavefront aberrations are rejected at the first order, residual effects are expected due to the expansion of the cloud during the interferometer sequence. Additionally, other atomic interferometers may have similar configuration to the one considered here, \textit{i.e.}, where due to expansion, the size of the atomic cloud reaches the size of the laser beam. This is all the more true, on the one hand, the power of the laser is limited, particularly in the context of space missions \cite{levequeCarioqaDefinitionQuantumPathfinder2022}, and its size must therefore be restricted in order to maintain sufficient intensity and, on the other hand, in order to improve the sensitivity of the experiment, a larger interferometer surface area and therefore a longer evolution time are sought. Finally, since laser beam distortions are today one of the main factors limiting the accuracy of existing experiments \cite{karcherImprovingAccuracyAtom2018,  LyuCompactGraviGradio2022, gaudout2025probingspatialdistributionkvectors}, it is important to be able to model and measure their impact in order to improve the error budget of current and future experiments \cite{levequeCarioqaDefinitionQuantumPathfinder2022, balaz2025longbaselineatominterferometry}.

%%%%%%%%%%%%%%%%%%%%%%%%%%%%%%%%%%%%%%%%%%%%%%%%%%%%%%%%%%%%%%%%
%%%%%%%%%%%%%%%%%%%%%%%%%%  Appendix  %%%%%%%%%%%%%%%%%%%%%%%%%%
\section*{Appendix: Calculations with contrast weights}\label{Appendix}
The contrast of the interferometer for a single atom is assumed to be of the form
\begin{equation}
    c = f \cdot \sin{\p{\mathcal{A} e^{-2\frac{\rho^2\p{t_3}}{w_0^2}} }} 
\end{equation}
$\mathcal{A}$ is the area of the third and last pulse of the interferometer, typically $\pi/2$, and $f$ is a function that does not depend on the transverse coordinates of the atom. The average contrast~\cite{hypergeometric} of the interferometer is then obtained by integrating this expression over the atomic cloud distribution
\begin{equation}\label{Eq:ContrastLoss}
    \begin{split}
        C &= \frac{1}{4\pi^2 \sigma_\rho^2\negthinspace\p{0} \sigma_v^2} \int e^{-\frac{ x^2 + y^2 }{2 \sigma_\rho^2\negthinspace\p{0}}} e^{-\frac{ v_x^2 + v_y^2 }{2 \sigma_v^2}} \cdot  c \cdot dx \thinspace dy \thinspace dv_x \thinspace dv_y \cdot \eta \negthinspace\p{z, v_z} dz \thinspace dv_z \\
        &= C_z \cdot \sum_{n=0}^{+\infty} \frac{\p{-1}^n \mathcal{A}^{2n+1}}{\p{2n+1}!} \cdot \frac{1}{1 + \p{2n+1} \p{\frac{2 \sigma_\rho\negthinspace\p{t_3}}{w_0}}^2 } \equiv C_z \cdot C_{\perp}.
    \end{split}
\end{equation}
$C_z$ is the integration of the function $f$ over the longitudinal distributions. The phase biases are calculated using expression~\eqref{Eq:PhaseShiftGen}, assuming that the wavefront aberrations depend only on the transverse positions of the atoms, which is valid as long as the size of the cloud along the optical axis is much smaller than the typical propagation length $l_z$ of the aberrations \cite{pagotInfluenceOpticalAberrations2025}.

The contribution of the curvature of a Gaussian beam wavefront $\delta \phi\left[\rho\negthinspace\p{t}\right] = \beta \, \rho^2\negthinspace\p{t}$ is
\begin{equation}\label{Eq:GaussianCurvLoss}
    \begin{split}
        \delta \phi &= \frac{\beta \, C_z}{4\pi^2 \sigma_\rho^2\negthinspace\p{0} \sigma_v^2C} \int e^{-\frac{ x^2 + y^2 }{2 \sigma_\rho^2\negthinspace\p{0} }} e^{-\frac{ v_x^2 + v_y^2 }{2 \sigma_v^2}} \cdot \rho^2\negthinspace\p{t} \sin{\p{\mathcal{A} e^{-2\frac{\rho^2\p{t_3}}{w_0^2}} }} \cdot dx \thinspace dy \thinspace dv_x \thinspace dv_y \\
        &= \frac{2 \beta \, \sigma_\rho^2\negthinspace\p{t}}{C_\perp} \sum_{n=0}^{+\infty} \frac{\p{-1}^n \mathcal{A}^{2n+1}}{\p{2n+1}!} \cdot \frac{\p{ 1 + \p{2n+1} \p{\frac{2 \sigma_\rho\negthinspace\p{0} \sigma_v (t_3 - t)}{w_0 \sigma_\rho\negthinspace\p{t}}}^2 } }{ \left[1 + \p{2n+1} \p{\frac{2 \sigma_\rho\negthinspace\p{t_3}}{w_0}}^2 \right]^2 }.
        %
        %\frac{2\p{ \sigma_\rho^2\negthinspace\p{t} + \p{2n+1} \p{\frac{2\sigma_0 \sigma_v (t_3 - t)}{w_0}}^2 } }{ \left[1 + \p{2n+1} \p{\frac{2 \sigma_\rho\negthinspace\p{t_3}}{w_0}}^2 \right]^2 }.
    \end{split}
\end{equation}
To the extent that the transverse selection is negligible $\rho\negthinspace\p{t_3} \ll w_0$, the expression converges toward $2 \beta \sigma_\rho^2\negthinspace\p{t}$, which is obtained with uniform contrast \cite{cervantesEffectAperture2024}.

The contribution of a wavefront aberration described by a Zernike polynomial, as in \cite{pagotInfluenceOpticalAberrations2025}, assuming that the Zernike polynomial defined on a disk of radius $R$ can be replaced by a Bessel function: $\delta \phi\left[\rho\p{t}\right] = \beta \, Z_n^0\p{ \frac{\rho\negthinspace\p{t}}{R} }\approx \beta \, J_0\p{ \frac{\rho\p{t}}{l_{xy}} }$ with $l_{xy} = R/(n+1)$, is 
\begin{equation}\label{Eq:ZernikeLoss}
    \begin{split}
        \delta \phi &=  \frac{ C_z }{4\pi^2 \sigma_\rho^2\negthinspace\p{0} \sigma_v^2 C} \int e^{-\frac{ x^2 + y^2 }{2 \sigma_\rho^2\negthinspace\p{0}}} e^{-\frac{ v_x^2 + v_y^2 }{2 \sigma_v^2}} \cdot \beta  \, J_0\p{ \frac{\rho\negthinspace\p{t}}{l_{xy}} } \sin{\p{\mathcal{A} e^{-2\frac{\rho^2\p{t_3}}{w_0^2}} }} \cdot dx \thinspace dy \thinspace dv_x \thinspace dv_y \\
        &= \frac{\beta}{C_\perp} \sum_{n=0}^{+\infty} \frac{\p{-1}^n \mathcal{A}^{2n+1}}{\p{2n+1}!} \cdot \frac{\exp{\p{\displaystyle -\frac{1}{2}\p{\frac{\sigma_\rho\negthinspace\p{t}}{l_{xy}}}^2 \cdot \frac{1 + \p{2n+1} \p{\frac{2 \sigma_\rho\negthinspace\p{0} \sigma_v\p{t_3 - t}}{w_0 \sigma_\rho\negthinspace\p{t}}}^2}{1 + \p{2n+1} \p{\frac{2 \sigma_\rho\negthinspace\p{t_3}}{w_0}}^2} } }}{1 + \p{2n+1} \p{\frac{2 \sigma_\rho\negthinspace\p{t_3}}{w_0}}^2 } \equiv \beta \, \mathcal{S}\negthinspace\p{t}.
    \end{split}
\end{equation}
Within the uniform contrast limit $\rho\negthinspace\p{t_3} \ll w_0$, the expression converges to $\beta \, e^{-\frac{1}{2} \p{\frac{\sigma_\rho\negthinspace\p{t}}{l_{xy}}}^2 }$, which is the one obtained in~\cite{pagotInfluenceOpticalAberrations2025}. $\beta$ contains the dependence on the amplitude of the aberration and on the ratio between the position along the optical axis and the typical propagation length of the Zernike aberration $l_z = 2 k l_{xy}^2$.

In the two previous expressions~\eqref{Eq:GaussianCurvLoss} and \eqref{Eq:ZernikeLoss}, the square of the cloud size is multiplied by the same correcting term
\begin{equation}\label{Eq:CorrectingFct}
    \begin{split}
        \sigma_\rho^2\negthinspace\p{t_i} \thinspace \cdot \thinspace & \frac{1 + \p{2n+1} \p{ \frac{2\sigma_\rho\negthinspace\p{t_3}}{w_0} }^2 \cdot \  \p{ \frac{\sigma_\rho\negthinspace\p{0}\sigma_v \p{t_3 - t_i}}{\sigma_\rho\negthinspace\p{t_i} \sigma_\rho\negthinspace\p{t_3} } }^2 }{  1 + \p{2n+1} \p{ \frac{2\sigma_\rho\negthinspace\p{t_3}}{w_0} }^2 }  \\
        & \hspace{0.5cm} = \sigma_\rho^2\negthinspace\p{t_i} \thinspace \cdot\p{ 1 - \frac{ \p{2n+1} \p{ \frac{2\sigma_\rho\negthinspace\p{t_3}}{w_0} }^2 \cdot \   \p{ \frac{\sigma^2_\rho\negthinspace\p{0} + \sigma_v^2 t_i t_3}{ \sigma_\rho\negthinspace\p{t_i} \sigma_\rho\negthinspace\p{t_3} } }^2 }{  1 + \p{2n+1} \p{ \frac{2\sigma_\rho\negthinspace\p{t_3}}{w_0} }^2 } }.
    \end{split}
\end{equation}
This effective reduction of the cloud transverse size corresponds to the fact that atoms with higher transverse velocities participate less to the interference process since they are on the edge of the laser beam during the final recombination pulse, where the atom-light coupling is lower. In particular, if the initial size of the cloud can be neglected with respect to the thermal expansion of the cloud, expression~\eqref{Eq:CorrectingFct} can be simplified as
\begin{equation}
    \frac{\p{\sigma_v t_i}^2}{ 1 + \p{2n+1} \p{ \frac{2\sigma_v t_3}{w_0} }^2}.
\end{equation}
For example, considering the contribution of defocus $Z_2^0$ corresponding to $l_{xy} = R/3$, the exponential in expression~\eqref{Eq:ZernikeLoss} can be linearized supposing $\sigma_\rho \ll R$. For low temperature ($2 \sigma_v t_3 \ll w_0$) the contribution is linear with the temperature,  and for higher temperature  ($2 \sigma_v t_3 \approx w_0$) the effect is reduced as seen in Figure~3 of \cite{karcherImprovingAccuracyAtom2018}. The analytical formulas are derived supposing that contrast losses due to the cloud transverse size at the first and second pulses are negligible.

\begin{backmatter}
\bmsection{Funding} The authors acknowledge the support from a government grant managed by the Agence Nationale de la Recherche under the Plan France 2030 with the reference “ANR-22-PETQ-0005” (project QAFCA).%, and the Agence Nationale de la Recherche for its financial support of the TONICS project "ANR-21-CE47-0017".% This work has been supported by R\' egion Ile-de-France in the framework of DIM SIRTEQ. %Ministère de l'Enseignement Supérieur et de la Recherche.

\bmsection{Acknowledgment}
The authors thank Arnaud Landragin for fruitful discussions. Optical elements in Figure~\ref{Fig1:Gradiometer} were taken from ComponentLibrary by Alexander Franzen, licensed under a Creative Commons Attribution-NonCommercial 3.0 Unported License.

\bmsection{Disclosures}
The authors declare no conflicts of interest.

\bmsection{Data availability} Data underlying the results presented in this paper are not publicly available at this time but may be obtained from the authors upon reasonable request.

\end{backmatter}

%%%%%%%%%% If using BibTeX:
\bibliography{sample}

\begin{thebibliography}{10}
\bibitem{series_n5}
The series comes from the development of the sinus in the contrast expression. The truncation of the sine series at $n=5$ is an $11^\text{th}$-order polynomial, and its difference with the sine function on $\left[0,\pi/2\right]$ is less than $10^{-7}$.
\bibitem{uncertainty_variance}
The variance estimator of $\left(X_i\right)_{1\leq i \leq n}$ is chosen to be $\overline{V}_n = \frac{1}{n} \sum_i \left( X_i - \overline{X}_n \right)^2$, with $\overline{X}_n=\frac{1}{n} \sum_i X_i$ the mean estimator. Statistical uncertainty is defined as the square root of the mean squared error (MSE) of the estimator. For the mean estimator: $\text{MSE}\left( \overline{X}_n \right) = \sigma^2\left(X\right)/n$. For the variance estimator: $\text{MSE}\left( \overline{V}_n \right) = \p{\mu^{(4)}\left(X\right) - \sigma\left(X\right)^4 }/ n + \mathcal{O}\p{n^{-2}}$, with $\mu^{(4)}\left(X\right) = \mathbb{E}\left[ \left( X - \mathbb{E}\left[X\right] \right)^4 \right]$ is the $4^\text{th}$ centered moment \cite{stats}. The standard deviation estimator is defined as $\overline{\sigma}_n = \sqrt{\overline{V}_n}$, with the associated uncertainty $\varepsilon \p{\overline{\sigma}_n} = \varepsilon \p{\overline{V}_n}/\p{2 \overline{\sigma}_n}$.
\bibitem{Zernike_similar}
The Zernike polynomials $Z_n^{|m|}$ and $Z_n^{-|m|}$ differ only by a rotation. Thus, simulations are performed for only one of the two polynomials.
\bibitem{Mirror_decomposition}
The mirror surfaces were decomposed over the first $2\thinspace500$ Zernike polynomials in the ISO-$14999$ indexing scheme. Analytical estimates of the biases were calculated using Zernike polynomials $Z_n^m$ with azimuthal indices $|m|\in\left\{ 0, 1, 2\right\}$, as adding those with $|m|=3$ results in changes smaller than the error indicated in parentheses in Table~\ref{Tab1:GradioRealMir}.
\bibitem{hypergeometric}
The series in equation~\eqref{Eq:ContrastLoss} is a hypergeometric function $\negthinspace _1F_2$.
\end{thebibliography}

\end{document}